\documentstyle[twocolumn,prl,aps,floats]{revtex}

\begin{document}
\input{psfig.sty}

\draft

\title{Calculation of the ultracold neutron upscattering 
rate from para-deuterium bound in a lattice}

\author{C.-Y. Liu$^1$, A.R. Young$^1$, and S. K. Lamoreaux$^2$}

\address{1.  Princeton University, Physics Dept., Princeton, N.J.\\
2. University of California, Los Alamos National Laboratory,
Physics Division P-23, Los Alamos, NM 87545}

\date{\today}

\maketitle

\begin{abstract}

The ultracold neutron (UCN) upscattering rate from para-deuterium 
contaminated solid D$_2$ near $T=0$ K
is calculated and
found to be  proportional
to the atomic fraction of para contaminant 
($x_{para}$) as $\tau_{up}^{-1}\approx 700\ x_{para} \  {\rm sec^{-1}}$.  
With $x_{para}=0.33$, corresponding to room temperature
equilibrium,
the upscatter time is 4.4 msec.  The implication is
that the para-deuterium contaminant is the temperature-independent
energy source responsible for the observed upscattering 
rate in the Los Alamos spallation driven UCN source.
\end{abstract}


\section{Introduction}

The possibility to use low neutron absorption cold materials,
such as solid deuterium, as a moderated source of ultracold neutrons (UCN)
has been long discussed \cite{shap72}.  However, as pointed out by
Golub and Pendlebury \cite{golpen}, one can in fact attain a much
higher UCN density
than implied by the thermal equilibrium arguments given in \cite{shap72};
if we consider UCN produced by inelastic down-scattering 
(in energy) of cold or
thermal neutrons, the
reverse process can be relatively suppressed.  In the case of
moderator with two energy levels, 
a ground state and an excited state separated by an energy
$\Delta$, the upscatter rate of a UCN with
energy $E_{ucn}$ and the downscatter rates of a neutron with
energy $E_{ucn}+\Delta$
rates are related according to microscopic reversibility:
\begin{equation}
\sigma_{up}=\sigma_{down} e^{-\Delta/kT}.
\end{equation}
The implication of this result is that, at very low temperatures,
UCN can be accumulated in a material until the rate of production $P$
equals the rate of loss, $1/\tau$,
due to nuclear absorption or leakage from a storage
volume that surrounds the converter material; the density
in equilibrium is
\begin{equation}\label{prod}
\rho=P\tau
\end{equation}
This is the so-called
``superthermal'' UCN production process, first described by
Golub and Pendlebury in relation
to thier proposed superfluid $^4$He UCN source.\cite{golpen,ucnbook}  
The theory of the superthermal
process for superfluid $^4$He
has been now very well established, both by observation
of upscattered UCN from a storage volume \cite{golup}, 
and by direct measurement
of the UCN density in a nearly losses superfluid $^4$He filled
magnetic trap.\cite{doyle}

The possibility of the use of other materials of low neutron absorption
cross section as a superthermal source
was first investigated by Golub et al. \cite{yumalgol} after
Mr. Kilivington observed the production of UCN in a prototype
superfluid $^4$He source, operated at the Institut Laue-Langevin,
even when it was empty of $^4$He; the explanation by Golub
was that there was
a film of dirt on the inner surface of the UCN
converter/storage vessel and this dirt was operating as a superthermal
source.
This led to the proposal of the ``thin film superthermal source'': Unlike the
case of $^4$He, all other materials have some neutron absorption cross
section.  For materials like solid D$_2$, the absorption lifetime is 
120 ms, compared to the effective lifetime of about 880 sec of
a UCN in superfluid $^4$He at low temperature, in which case $\beta$-decay
is the ultimate limiting loss mechanism.  The idea of the thin film
source is that the UCN density in a storage volume of
volume $V$ that contains
some converter material is independent of the volume of the
converter material  $V_c$ because the total production rate is proportional
to $V_c$, while the net effective lifetime $\tau$ is proportional to $1/V_c$
(in the limit where upscattering and absorption in the converter material
dominates all other loss mechanisms),
and $V_c$ cancels in Eq. \ref{prod}.  $V_c$ determines the time constant
for filling the volume $V$ to an equilibrium density.

The figure of merit for a superthermal source is given by $P\tau$
for the converter material.  Superfluid $^4$He is at least an order
of magnitude better than any other material; in this case,
even though $P$ is relatively low, the long intrinsic lifetime
more than makes up the difference in production
rate compared to other materials.  However, there are
situations where use of a less efficient material is warranted.
For example, a thin film source might be better suited near the
core of a nuclear reactor because less converter material is required
to achieve the intrinsic maximum UCN density.  

Another situation where a material with a large $P$ but short $\tau$
was first pointed out by Pokotilovski \cite{poko} in connection
with pulsed neutron sources, such as spallation sources.  The
idea is that the converter material is connected to a UCN storage
volume only for a short time during and after the neutron pulse,
after which a valve is closed, separating the conversion and
storage regions.  In this case, one can enjoy a relatively large $P$
together with an net effectively large $\tau$.  
Serebrov elaborated upon this idea
with the proposal of a ``UCN Factory'';  following this lead,
we developed a prototype UCN source that turns out to follow the
original Pokotilovski idea more closely, but ours incorporates a
cold neutron flux trap, and vertical UCN extraction from the
solid D$_2$ converter to counteract the solid deuterium positive
potential step.

In our recent studies of solid deuterium as a spallation driven
generator of ultracold neutrons, the apparent production rate
is roughly an order of magnitude less than expected, and
there is a lack of temperature dependence on the net production
rate, contrary to the prediction
by Golub et al. \cite{yumalgol,ucnbook}  The reason
for the reduction appears primarily due to a loss of UCN
before they can exit
the solid deuterium.   
By elimination of various loss  sources within the material that
can account for the temperature independent loss, we came
to the conclusion that the UCN loss is likely due to
upscattering  from  para-deuterium contaminant
with solid (ortho-) deuterium converter material; in addition,
anecdotal evidence for this was offered by Serebrov. 
The following calculation estimates the upscatter rate and
was motivation for incorporating a para/ortho converter\cite{bazh}.
The preliminary results indicate the following calculation is correct;
the data and analysis will be presented in a full publication by
the collaboration.  We felt it important, however, to make available
our initial estimate of the upscatter rate due to the para contaminant.

\section{Properties of D$_2$ and assumptions in the calculation}

The deuterium molecule exists in the ground state as a total
nuclear spin $S=0$ and rotational angular momentum $J=0$,
and is referred to as $ortho-$deuterium.

The first excited state has $S=1$ and $J=1$, and the energy relative
to the ground state is
\begin{equation}
E_0=7\ {\rm meV}.
\end{equation}
We refer to this state as $para-$deuterium.

Of course, at high temperature, the $S,J$ states are occupied
according to their degeneracy and Boltzmann factor; just considering
the nuclear states, the ortho-states $(S=0,2)$ and para-states $(S=1)$
are occupied at high temperatures according to their spin degeneracies:
\begin{equation}
x_{para}={3\over 1+3+5}=0.33;\ \ \ x_{ortho}=
{1+5\over1+3+5}=0.67
\end{equation} 
because the degeneracy is given by $2S+1$.

The rotational states are coupled strongly to the lattice and we can
assume the $J$ populations follow the Boltzmann distribution with
rapid equilibration.  On the other hand, the transitions between 
ortho and para is very slow under normal conditions, and we can assume
that the room temperature composition persists as the D$_2$ sample
is cooled and solidified, and is rather permanent in our experiment.

We can assume the sample is at absolute zero; we therefore can
neglect upscattering from the vibrations of the lattice which represents
a different upscatter mechanism, roughly independent of upscattering
from the para contaminant.  Furthermore, we can make an approximation
that the molecules are fixed in the lattice and neglect recoil
effects; this approximation is valid because a) $E_0<\theta_D$, the
Debye temperature, b) the momentum transfer is in the form of
angular momentum from
the molecular rotation to the outgoing wave, c) the UCN energy
$E_i<<\theta_D$, making direct excitations of the lattice unlikely.
Finally, we assume that all the para molecules are in the $J=1$
rotational state, while the ortho molecules are in the $J=0$ state.
This final assumption is valid because the rotational levels, unlike
the nuclear spin state, are closely coupled to the thermal bath, as
mentioned above.

Normally, we assume that $P-$wave scattering is suppressed for cold
and ultracold neutrons.  However, in this case, the scattering is
from a molecule; the separation between the deuterium atoms in the
molecule is comparable to the wavelength of an outgoing upscattered
UCN (with final energy 7 meV) and we do not have the usual $P-$wave
scattering suppression found in nuclear scattering from a single
isolated nucleus.

\section{Upscatter rate}

Much of Young and Koppel (YK) \cite{yandk} is irrelevant to this calculation
and in addition is not entirely transparent.  Our problem is
simple enough to begin from first, very basic, principles.  We will
see than only a single matrix element from YK is required for our
calculation, as subject to the above constraints and approximations.

Starting with Fermi's Golden Rule for transitions to a continuous 
state (see \cite{ll}, Sec. 43)
\begin{equation}
dw_{fi}=(2\pi/\hbar)|U_{fi}|^2\delta(E_f-E_i){\rm d} \nu_f
\end{equation}
where $dw_{fi}$ is the transition rate
between the levels $f,i$, ${\rm d} \nu_f$ is the density of
final states, and $U_{fi}$ is the matrix element of the Hamiltonian between
the initial and final states,
\begin{equation}
U(\vec r)=-{2\pi\hbar^2\over m}\left(a_1\delta(\vec r-\vec r_1)+a_2\delta
(\vec r-\vec r_2)\right]
\end{equation}
where $1,2$ label the two atoms of the molecule, $m$ is the neutron
mass, and $a_{1,2}$ are
the scattering lengths, chosen to represent the ortho- or para- states 
as required.

First, let us take the incoming and outgoing neutron states as
plane waves (as did YK).  The incident (UCN) wave must represent
the UCN density $\rho_u$ in a normalization volume $V$.  Thus, 
\begin{equation}
\psi_{i,n}={1\over \sqrt{V}}{\sqrt{\rho_u V}}e^{i\vec k\cdot \vec r}=
\sqrt{\rho}
\psi_i
\end{equation}
where the factor $1/\sqrt{V}$ normalizes the incident wave to the
volume $V$, the second factor gives the total UCN $n=\rho_u V$ within
$V$, the subscript $i,n$ means incident wave, normalized, and
$\psi_i$ conforms to the states used by YK.

The outgoing wave can be written
\begin{equation}
\psi_{f,n}={1\over \sqrt{V}}e^{-i \vec k_f\cdot r}=\psi_f/\sqrt{V}.
\end{equation}

The density of final states $\vec k_f$ is
\begin{equation}
{\rm d}\nu_f={V {\rm d}\vec k_f\over (2\pi)^3}=
{V k_f^2{\rm d} k_f {\rm d}\Omega\over (2\pi)^3}.
\end{equation} 

In Fermi's Golden rule, the $\delta$-function ensures energy
conservation.  In the present case, 
\begin{equation}
E_f-E_i=\hbar^2k_f^2/2m-\hbar^2k_i^2/2m-E_0
\end{equation}
and we have $\delta(f(k_f))$.  Recall the well-known relationship,
\[
\delta(f(x))={1\over |f'(x_0)|}\delta(x-x_0)
\]
where $x_0$ is the zero of $f(x)$.  A further assumption $\vec k_i=E_i=0$
is valid for UCN in this case.  Then the momentum transfer $\vec \kappa
=\vec k_f$, in YK's notation.

Therefore, the upscatter rate is
\begin{eqnarray*}
dw_{fi}&=&{2\pi \over \hbar} \left({2\pi\hbar^2\over m}\right)^2 \\
&\times&|\langle \psi_f,J'=0,S'=1|\sum a_i| \psi_i J=1,S=0\rangle|^2 
{m\over \hbar^2 k_f} \\
&\times& \delta(k_f-\sqrt{2mE_0/\hbar^2}){Vk_f^2dk_f\over (2\pi)^3}
\end{eqnarray*}
which must be integrated over $k_f$; this is easily done, and
we write the matrix element $|\sum a_i|^2$, yielding
a total rate, taking $Vx_{para}\rho_{D_2}$ scattering molecules in $V$,
\begin{equation}
w_{fi,tot}=4\pi V x_{para}\rho_{D_2} \rho_{ucn} {\hbar\over m} k_0 |\sum a_i|^2
\end{equation}
where
\begin{equation}
k_0=\sqrt{2mE_0\hbar^2}=1.83\times 10^8 {\rm cm}^{-1}
\end{equation}
and
\begin{equation}
{\hbar\over m}k_0=v_0=1.2\times 10^{5} {\rm cm/s}.
\end{equation}
We further note that $w_{fi,tot}/V=\dot \rho_{ucn}$ so we have a loss
rate of
\begin{equation}
\Gamma=4\pi x_{para}\rho_{D_2} v_0 |\sum a_i|^2
\end{equation}

Thankfully, YK have calculated the matrix elements.  We only need
\begin{eqnarray}
|\langle k_f, J' &=&0,S'=1|\sum a_i| k_i=0,J=1,S=0\rangle|^2=\\
&=&{3\over 4}
a_{inc}^2C^2(1011:00)|A_{01}|^2\\ & = & {3\over 4} a_{inc}^2 |A_{01}|^2
\end{eqnarray}
from Eq. (A5) of KY and Appendix B. Also, ($a=.742\AA$ so $k_0a/2=0.68$) 
\begin{equation}\label{anl}
|A_{01}^2|=|2ij_1(k_0 a/2)|^2=4\times j_1(k_0 a/2)^2=4\times(.21)^2=.18
\end{equation}
so
\begin{equation}
|\sum a_i|^2=0.13 a_{inc}^2.
\end{equation}
Taking $a_{inc}=4.04\times 10^{-13}$ cm gives
\begin{equation}\label{res}
\Gamma=3.2\times 10^{-20} x_{para}\rho_{D_2} {\rm s^{-1}}
\end{equation}
and, with  $x_{para}=0.33$ corresponding to room temperature
equilibrium, and $\rho_{D_2}=3\times 10^{22}/{\rm cm^3}$ gives
\begin{equation}
\Gamma=316\ {\rm sec^{-1}};\ \ \ \ \tau=3.2\ {\rm ms}.
\end{equation}

Finally, 
zero-point motion of the lattice leads to a smearing-out
of the final state, and a reduction in the cross section.  This
effect is given by the Debye-Waller factor, which for solid deuterium
is a factor of 0.72 for a momentum transfer corresponding to 7 meV.
\cite{sold2}
The net upscatter rate and lifetime is thus
\begin{equation}
\Gamma=221\ {\rm sec^{-1}};\ \ \ \ \tau=4.4\ {\rm msec}.
\end{equation}

\section{Discussion}

The important implication of this calculation is that the upscatter
rate from ortho-deuterium is independent of temperature and persists
at absolute zero.  The agreement between this simplified calculation
and experiment is satisfactory and appears to have identified
a principal limitation to the present Los Alamos UCN source density.

The surprising result is that the $P-$wave scattering is significant;
the outgoing plane wave state carries away the rotational angular
momentum of the molecule.  This is possible because the impact
parameter, corresponding to the molecular size, is comparable to
the outgoing neutron wavelength, as is illustrated in 
Eq. \ref{anl}.  

\section{Conclusion}

Contamination of a solid deuterium UCN source by the para-molecular
state will severely limit the attainable UCN density.  Given the stability
of the para-deuterium state (0.02\%/hr conversion rate in the solid)
it is clear that the gas must be converted before freezing.  The
details of the Los Alamos converter will be published, but it follows
standard techniques.\cite{bazh}  Our preliminary results support the above
calculated rate.

Our result has serious implication for solid deuterium sources operated
close to the core of a nuclear reactor.  In this situation, the
gamma flux will be sufficient to drive molecular excitations, and
the ortho-para ratio will likely be near the high temperature equilibrium.
This implies that solid deuterium might be useless in such an 
environment, although the presence of free charge in the lattice
might catalyze the conversion.  We are presently investigating
these and other effects expected in a high radiation environment.
We further point out that the radiation levels in our proposed
spallation driven source are many orders of magnitude less than
a reactor so we expect no problems in this regard.

A more complete calculation of the upscattering rate and related 
issues is  in
progress by the authors.

\end{document}